\documentstyle[rotating,longtable,epsfig,cite_extras]{mn}
\newcommand{\rg}{radio galaxy}
\newcommand{\rgs}{radio galaxies}

\newcommand{\hirs}{high--redshift}

\newcommand{\ackpparc}{REJS acknowledges the support of a PPARC studentship.  }
\newcommand{\ackads}{This research has made use of NASA's Astrophysics Data 
	System Abstract Service.  }

\newcommand{\etal}{{et al.}}

\newcommand{\ltabout}{\stackrel{<}{_{\sim}}}
\newcommand{\ga}{\stackrel{>}{_{\sim}}}
\newcommand{\la}{\stackrel{<}{_{\sim}}}
\newcommand{\sixc}{6C\,0140$+$326}
\newcommand{\eightc}{8C\,1435$+$635}
\newcommand{\btwo}{B2\,0902$+$343}
\newcommand{\fourc}{4C\,41.17}

\def\myensuremath#1{\ifmmode#1\else$\relax#1$\fi}

\newcommand{\Msun}{\mbox{$M_\odot$}}

\newcommand{\steidU}{\mbox{${U_n}$}}
\newcommand{\steidG}{\mbox{${G}$}}
\newcommand{\steidR}{\mbox{$\mathcal{R}$}}

\newcommand{\HI}{H{\sc i}}
\newcommand{\HII}{H{\sc ii}}

\newcommand{\HeII}{He{\sc ii}}
\newcommand{\lya}{\nobreak\mbox{$\mathrm{Ly}\alpha$}}

\newcommand{\Ho}{\nobreak\mbox{$\mathrm{H}_0$}}

\newcommand{\prep}{\mbox{astro-ph/}}

\title{Two-colour photometric selection of high-redshift galaxies}

\author[Stevens \& Lacy ]
{Robin Stevens$^1$\footnotemark \& Mark Lacy$^{1,2,3}$ \\
$^1$ Astrophysics, Department of Physics, Keble Road, Oxford, OX1 3RH.
Electronic mail: \texttt{rejs@astro.ox.ac.uk}\\
$^2$ 
I.G.P.P., L-413 Lawrence Livermore National Laboratory, 7000 East Avenue,
Livermore CA 94550\\
$^3$Dept.\ of Physics, University of California, Davis}

\begin{document}

\maketitle

\begin{abstract}

In this paper we describe a set of models to predict the colours of
galaxies over a wide range of redshifts.  We present example output from
the simulations, and discuss their application to the selection of galaxies
at high redshifts, particularly through identification of the Lyman break.
Additionally we consider the optimal choices of filters for selection at a
range of redshifts. 

An interface to a subset of the simulations has been made available on the
World Wide Web for the benefit of the community at the location
\texttt{http://www-astro.physics.ox.ac.uk/\~{ }rejs/research/galcols.html}.

\end{abstract}
\begin{keywords}
galaxies:$\>$formation -- galaxies:$\>$distances and redshifts -- galaxies:
fundamental parameters -- galaxies:$\>$photometry
\end{keywords}

\section{Introduction}

\footnotetext{Present address: Oxford University Computing Services, 
13 Banbury Road, Oxford, OX2 6NN}

In recent years, two-colour selection techniques have proved themselves to
be an invaluable method for identifying galaxies at high redshifts,
paricularly by means of chosing a filter set spanning the wavelengths
between \lya\ and the Lyman limit.  Most notable has been the work of
Steidel and collaborators (e.g.\ \pcite{steidspec96,steidg99}).  

As part of a project to identify radio-quiet companions to known \rgs\ in
the redshift range $3 \la z \la 4.5$, we required theoretical models to
determine the optimal selection criteria given our choice of filters.
This paper discusses modelling of the expected colours for galaxies and the
potential for efficiently distinguishing them from foreground stars and
galaxies through broad-band photometric techniques.
The results of our observations are described in a companion paper
\cite{rejspaper}.

In our searches for Lyman-break galaxies (hereafter LBGs), we initially
worked to the assumptions that their spectra will be approximately flat
above the rest-frame \lya\ wavelength, somewhat diminished between \lya\
and the Lyman limit, and have little or no flux below 912\AA.  Such
assumptions were used for our pilot study of a region of approximately one
square arcminute around the $z=3.80$ \rg\ \fourc, described in
\scite{mdlsr:4c}.

Unfortunately such na\"{\i}ve selection methods lend themselves to
significant contamination from other red objects such as galaxies at
moderate redshift (for instance where the 4000\AA\ break is redshifted to
the position of the Lyman break at the targetted redshift) or low-mass
main-sequence stars.  With the benefit of the models described in this
paper and of Hubble Space Telescope images of the region studied in our
pilot study, which have subsequently been made available through the HST
Archive, we conclude that the majority of the \hirs\ candidates previously
identified are in fact foreground contaminants.

For more efficient selection, a much more detailed analysis will be
required and the following points must be considered within the models:

\begin{enumerate}
\item Intrinsic spectra of starbursts
\item Attenuation due to intergalactic \HI\ absorption systems
\item Effects of dust extinction
\item Broad-band colours and potential for confusion with other objects
through photometric selection techniques
\item Apparent magnitudes of high-redshift star-forming galaxies
 for particular cosmologies, likely star-formation rates and
consequences for their observability
\end{enumerate}

Throughout this paper, we follow the convention established by Steidel
\etal\ \cite{steid1} of using the AB magnitude system of \cite{oke:abmags},
conversions for which are given in Table~\ref{tab:abmags}.
This has the great advantage of being a magnitude system in which the
colours of an unreddened star-forming galaxy between bandpasses sampling
the rest-frame ultraviolet continuum will be close to zero.

\begin{table}
\label{tab:abmags}
\begin{minipage}[c]{\linewidth}
\begin{center}
\begin{tabular}{lrr}
Filter & $\lambda_{\mathit{eff}}$  (\AA)& $m_{\mathrm{AB}}$(Vega) \\
\hline
\hline
  $U$ &   3664  &   0.780 \\
  $B$ &   4394  &  $-$0.105 \\
  $G$ &   4869  &  $-$0.098  \\
  $V$ &   5449  &   0.000 \\
  $R$ &   6396  &   0.186 \\
  $I$ &   8032  &   0.444  \\
\hline
\end{tabular}
\end{center}
\caption{Conversions between AB and Vega-based magnitudes of the principal
filters used in our models.
}
\end{minipage}
\end{table}

\subsection{Previous work}

Guhathakurta and co-workers \cite{gtm90} searched for LBGs in relatively
deep $UB_{J}RI$ images, but were unable to provide followup spectroscopic
evidence.  They estimated the strength of the break at the Lyman limit from
consideration of the UV spectra of O and B stars, with synthetic spectra
showing breaks of a factor in the range $2-5$ for a wide
variety of input initial mass functions (IMFs), even without considering
interstellar or intergalactic absorption by neutral hydrogen.  Candidate
LBGs are selected from their $U-B_{J}$ and $B_{J}-I$ colours.  Their
resulting selection region is somewhat conservative compared to later work,
as it does not take into account the possibility of a substantial drop in
flux below \lya.

Most of the work of Steidel and collaborators makes use of a non-standard
filter system, originally chosen for its suitability to a field containing
a QSO and damped Lyman $\alpha$ system at known redshifts \cite{steid1}.
Initially, selection is done taking into account intergalactic absorption
measurements from QSO spectra \cite{sargent89}, while later selection
criteria are revised to take into account the detailed analysis of
intergalactic attenuation by \scite{madau:cols95}, which we use ourselves
in \S~\ref{sec:attenuation}.  Subsequent spectroscopic observations show
that the selection process is highly effective with less conservative
selection criteria.  The lower bound in $\steidU -\steidG$ has been relaxed
from $1.5$ to $1.0$, with a resulting increase in surface density (to the
magnitude limit of $\steidR < 25.0$) from around 0.5 candidate objects
per arcmin$^2$ to around 1.25 per arcmin$^2$
\cite{steidelfrascati,adelberger98}.

\scite{madauhdf96} use a similar approach to that described here in order
to determine appropriate regions in two-colour space for the selection of
$U_{300}$ and $B_{450}$ dropouts in the northern HDF.  We develop this
approach to a wider range of filters, including those typical of those
found on ground-based optical CCD and NIR cameras, and to span as wide a
range of potential galaxy scenarios as possible.

\section{Simulated broad-band colours of galaxies}\label{sec:sims}

In order to devise efficient selection criteria for \hirs\ galaxies, it is
necessary to model the expected colours of a wide variety of different
types of galaxies for all redshifts at which one might possibly expect to
be able to see objects.

We base our models on the  {\sc p\'{e}gase} code of \scite{frv97}
(hereafter FRV); our thanks go to them for making the code publicly available.  
\footnote{A more recent version of the code has recently been announced
\cite{frv00}, 
but at present our models are built around the version one code.}
The FRV code produces synthetic galaxy spectra to a resolution of 10\AA\ 
from 220--8780\AA\ in the rest-frame\footnote{This resolution was higher
than that of the Bruzual-Charlot models available at the time
\cite{brch93}, an important factor when one wishes to shift the spectra to
high redshifts.}, and to 200\AA\ in the NIR.  Like most current techniques
for modelling the integrated spectrum of a population of stars, the FRV
code uses the isochrone technique \cite{chbr91}.  A library
of higher-resolution spectra is available, but this only covers the rest
frame optical and is thus of little use when the objects are redshifted;
besides which the higher resolution data is not overly crucial when dealing
with broad bandpasses. 

A substantial library of synthetic galaxy spectra using a wide
variety of different input parameters to the FRV code was built up.

\subsection{Variations of input parameters}

\subsubsection{Evolutionary tracks}

Two sets of stellar evolutionary tracks are made available with the FRV
code, namely ``Padova'' \cite{bressan93} and ``Geneva''
\cite{schaller92,charbonnel96}.  Post main sequence evolution, in
particular that along the asymptotic giant branch for the most massive
stars \cite{jimenez:ssp1}, proves the most difficult phase to trace in such
tracks. Primarily this affects the NIR luminosity, with only a slight
effect in the UV.

Another limitation is that the evolutionary tracks used are for solar
metallicity, and will therefore not be an accurate representation of those
in metal-poor regions such as may be found in the early universe.  The
version 2.0 FRV code attempts to redress this problem \cite{frv00}, and
will be incorporated into our models at a future date.  Early indications
are that metallicity significantly affects the slope of the UV
continuum, with bluer spectra at low metallicities.  Evidently this will
have a considerable effect upon the derived star formation rates.
Further work with the new models is required to investigate this more
thoroughly.

\subsubsection{Initial Mass Function}\label{sec:imf}

The FRV code by default incorporates the IMFs of \scite{salpeter55},
\scite{scalo86}, \scite{millerscalo79} and of Kroupa \cite{kroupa93}.  The
Salpeter is the simplest and still in widespread usage, but somewhat in
error at the extremes.  Later IMF models are an improvement in this
respect, being somewhat more realistic and with less abrupt cut-offs at
both ends, but are still limited by the the ability to observe the full
range of stellar masses only in the solar neighbourhood.  

\begin{figure}
\begin{center}
\epsfig{file=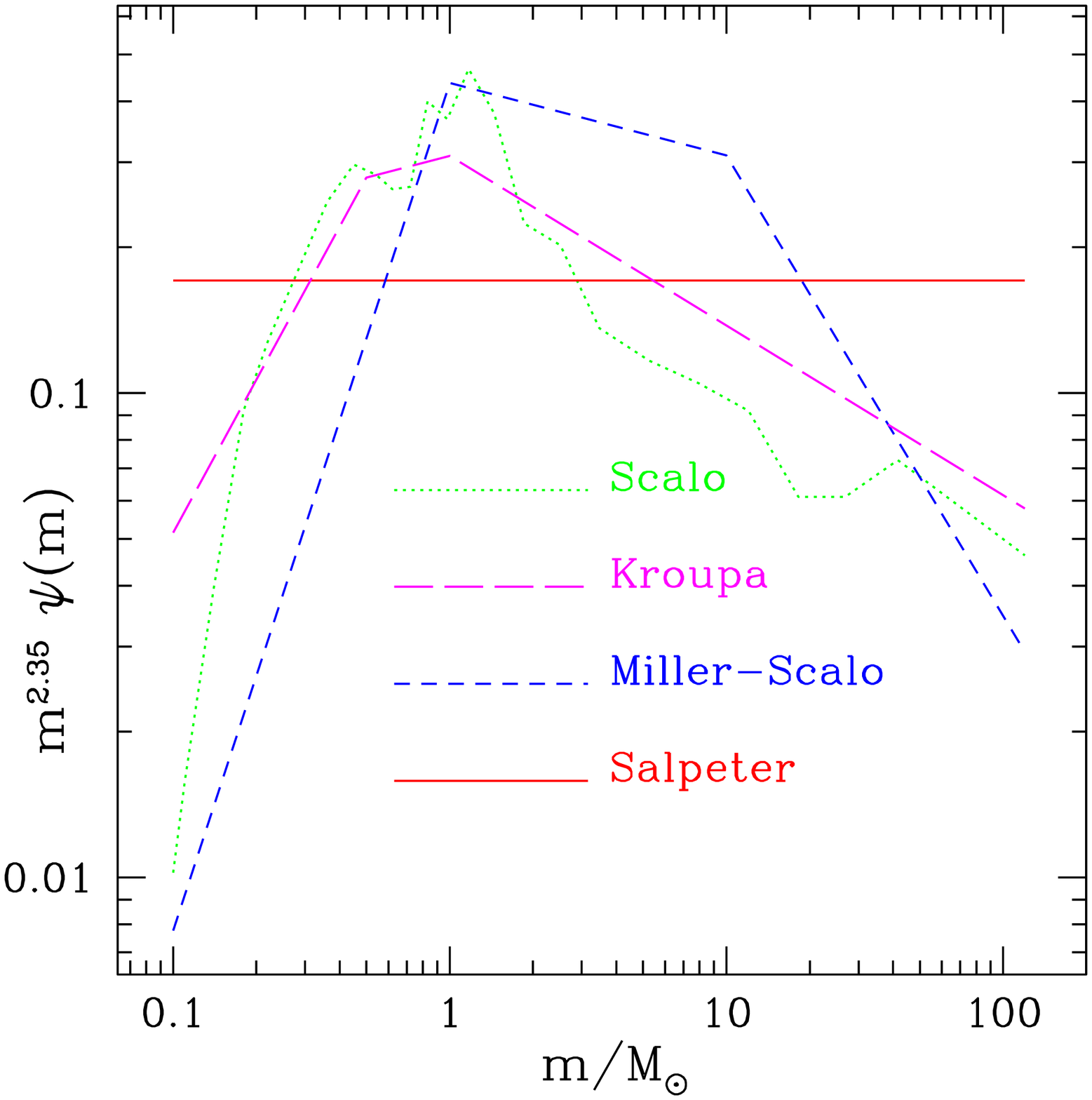, width=70mm, height=70mm}
\vspace{-4mm}
\end{center}
\caption\protect{Representations of the four initial mass functions
 used in the simulations.
For clarity, all have been multiplied up by a factor $m^{2.35}$; this gives
a constant slope for the Salpeter IMF.
\label{colfig:weighted-imf}
}
\end{figure}

The Kroupa IMF results from work on stars in the galactic disk which
resolves the apparent differences between observations made in the solar
neighbourhood and those derived from surveys to much greater distances. It
is probably to date the most realistic representation of the IMF in our own
galaxy and it is used as the default in our models.

Applying such empirically-derived IMFs to star-formation in the early
universe is a big assumption, given our limited understanding of the
physical processes involved. For example, low metallicity, different cloud
dynamics and the possible influence of AGN activity upon star-formation may
all affect the \hirs\ IMF.

The high-mass slope of the IMF is of particular interest to us since it
determines the numbers of O and B stars which dominate the UV luminosity in
star-forming galaxies.  Any estimate of star-formation rate based on the UV
luminosity is thus strongly dependent on the choice of IMF.  For simplicity
and ease of comparison, star formation rates are generally derived for a
Salpeter IMF, but one must bear in mind that with, for example, a Kroupa
IMF as opposed to a Salpeter distribution, the overall star-formation rate
may be a factor of $\sim2$ higher if derived from the UV luminosity.  The
effects of extinction due to dust will boost the SFR higher still.

\subsubsection{Nebular emission}

FRV use a default value for the fractional Lyman continuum absorption
$f=0.7$, in agreement with LMC \HII\ values \cite{dge92}, but for
completeness values from 0 to 1 were considered.  This would have
considerable effect on the strength of the Lyman break save for the fact
that intergalactic attenuation becomes the dominant effect at redshifts
sufficiently high for us to observe below the Lyman limit
(\S~\ref{sec:attenuation}).  However it does govern the strength of
emission lines, which, particularly in the case of \lya, can in principle
have considerable effect on the observed colours, though it is often
strongly suppressed.  Continuum emission is increased in the rest-frame
optical, but this will only have an effect on the infrared colours of
galaxies at high redshift.

\subsection{Star-formation model}

A variety of star-formation models are included in the FRV code, ranging
from instantaneous bursts to constant SFR scenarios.  Primarily the choice
affects how quickly the galaxy spectrum evolves from that of young 
starbursting regions to that of aged ellipticals.  We include in the
final models galaxy spectra covering a very wide range of star-formation
models and ages, subject to a conservative cut-off to eliminate galaxies at
ages unphysical for any reasonable cosmology given their redshift.  Thus
the models include more or less any conceivable galaxy type.

\subsection{Other parameters}

The models also incorporate variations of initial star formation rate,
governing the timescale over which stars are formed, and the fraction of
stellar ejecta available for further star formation.  This latter only
affects galaxies sufficiently aged for a substantial proportion of the
stars originally formed to go through the late stages of stellar evolution;
significant star formation resulting from stellar ejecta results in the
overall galaxy spectrum being a combination of the UV spectrum of a young
star-forming galaxy together with the optical and NIR spectrum expected of
an aged galaxy. 

Dust extinction is also considered, but rather than than use the FRV
extinction routine, we chose to use our own as described in
\S~\ref{sec:dustext}.

The spectra were then convolved with selected broad-band filter profiles
(those of the WHT optical and NIR filter sets, and the HST WFPC2 filters as
used in the HDF observations), scaled to the luminosity of a galaxy of
stellar mass $10^{10}$\Msun, and the AB magnitudes output. 

\begin{figure}
\begin{center}
\epsfig{file=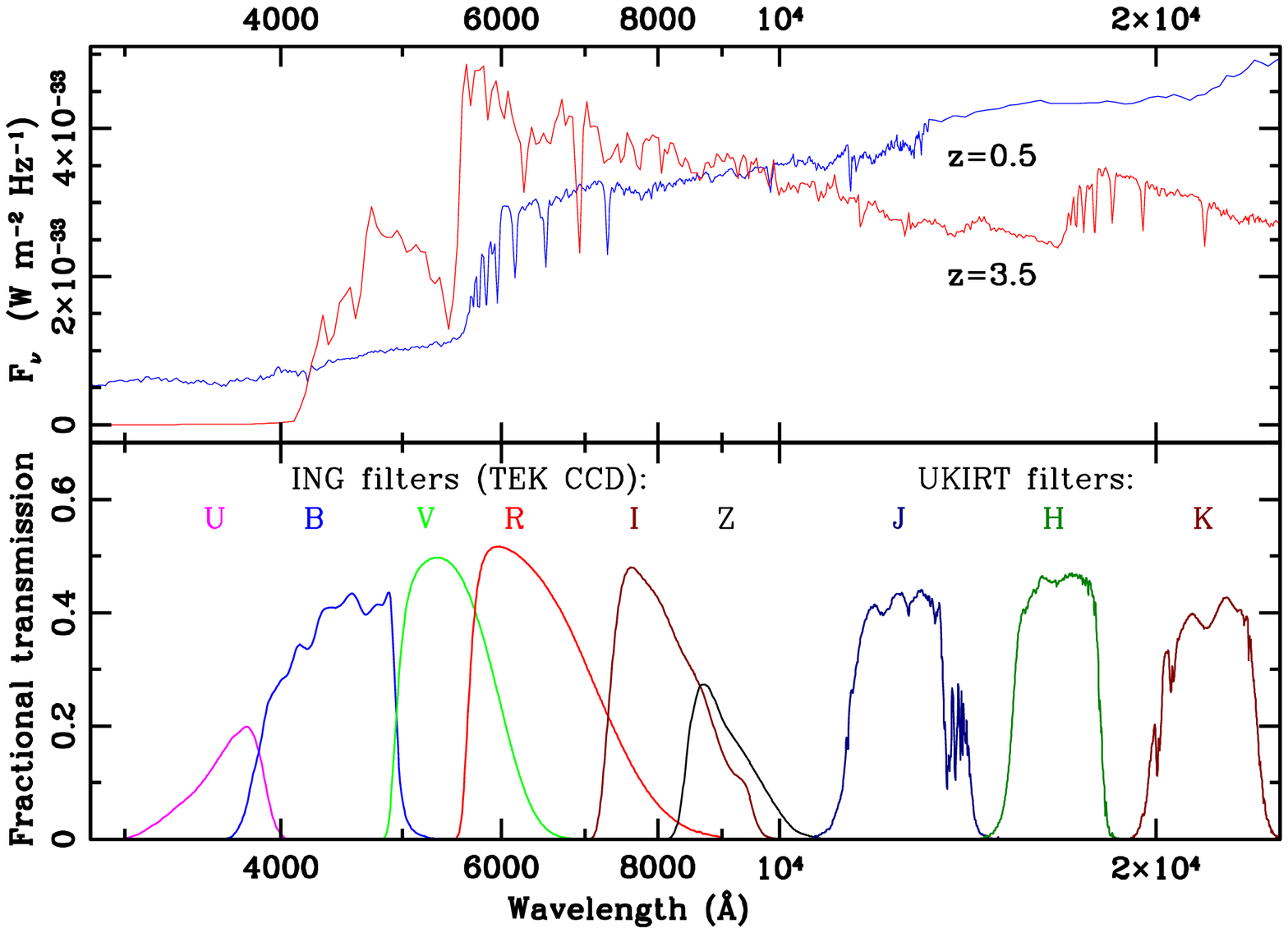, width=82mm, height=57mm}
\end{center}
\caption\protect{Model spectra in $f_\nu$ for a $z=3.5$ galaxy undergoing
star formation at a constant rate of 10\Msun\ yr$^{-1}$, and for an evolved
galaxy at $z=0.5$, typical of lower-redshift galaxies likely to have
similar colours to the targetted \hirs\ objects owing to the presence of
the 4000\AA\ break. Our selection criteria must therefore be able to
discriminate effectively between the two, using the much bluer continuum
colour of the high-redshift object relative to the low-redshift galaxy.  
The Lyman-$\alpha$ and Lyman-limit breaks are clearly visible in the
former; similar objects may possibly pass the selection criteria for both
$U$ and $B$ dropouts. 

Also shown are the response curves for various filters.  The optical
filters are those from the Isaac Newton Group telescopes on La Palma and
include those used in our observations (a similar set of filters plus the
$Z$ is being used for the wide field survey of Dalton \etal\ currently in
progress at the INT); all filter profiles have been convolved with the
response curve for the TEK CCD used in our observations.  
The $JHK$ filter profiles are taken from data available on the UKIRT WWW pages.

No scan was available for the $G$ filter (loaned by Richard McMahon); for
these models a bandpass of width 1000\AA\ centred at 4900\AA\ has been
assumed, as per the design specifications.
\label{colfig:examplespec}
}
\end{figure}

\section{Intergalactic attenuation by the Lyman-$\alpha$ forest and by
Lyman limit systems}\label{sec:attenuation}

To approximate intergalactic attenuation, the method of
\scite{madau:cols95} was followed.  Equations 12 and 13 of his paper
were used to correct for the Lyman-$\alpha$ forest.  To correct for
absorption of the Lyman continuum we made use of the approximate numerical
integration of his Equation 16, which he states is good to 5\%.  For our
purposes this is perfectly adequate and computationally inexpensive.  

Given the random distribution of the absorbers, significant deviations from
the mean attenuation are to be expected at any particular wavelength.  In
the Lyman-$\alpha$ forest region, this does not present any problem, since
the use of broad bandpasses means that one will be averaging over a large
number of absorbers.  

Madau's figure~2 shows the measured flux decrements for quasar spectra within
bandpasses of $\sim$100\AA\ in the rest-frame.  Few decrements differ from
the  estimated mean by more than about 0.1~mag.  In our case we can expect
an even greater averaging effect to act in our favour owing to the wider
bandpass of our filters. 

Below the Lyman limit, a substantial contribution to the attenuation will
come from the much rarer Lyman limit systems (LLS), for which the mean
number along any line of sight will be of order unity.  There is, however, a
significant probability that there will not be a single LLS along a
particular line of sight, about 10\% at the redshifts of interest.
While in the case of the \lya\ forest, the distribution of the number
densities of absorbers about the mean can be approximated by a Gaussian at
the redshift of interest, such an approximation cannot be made for the LLS. 

\begin{figure}
\begin{center}
\epsfig{file=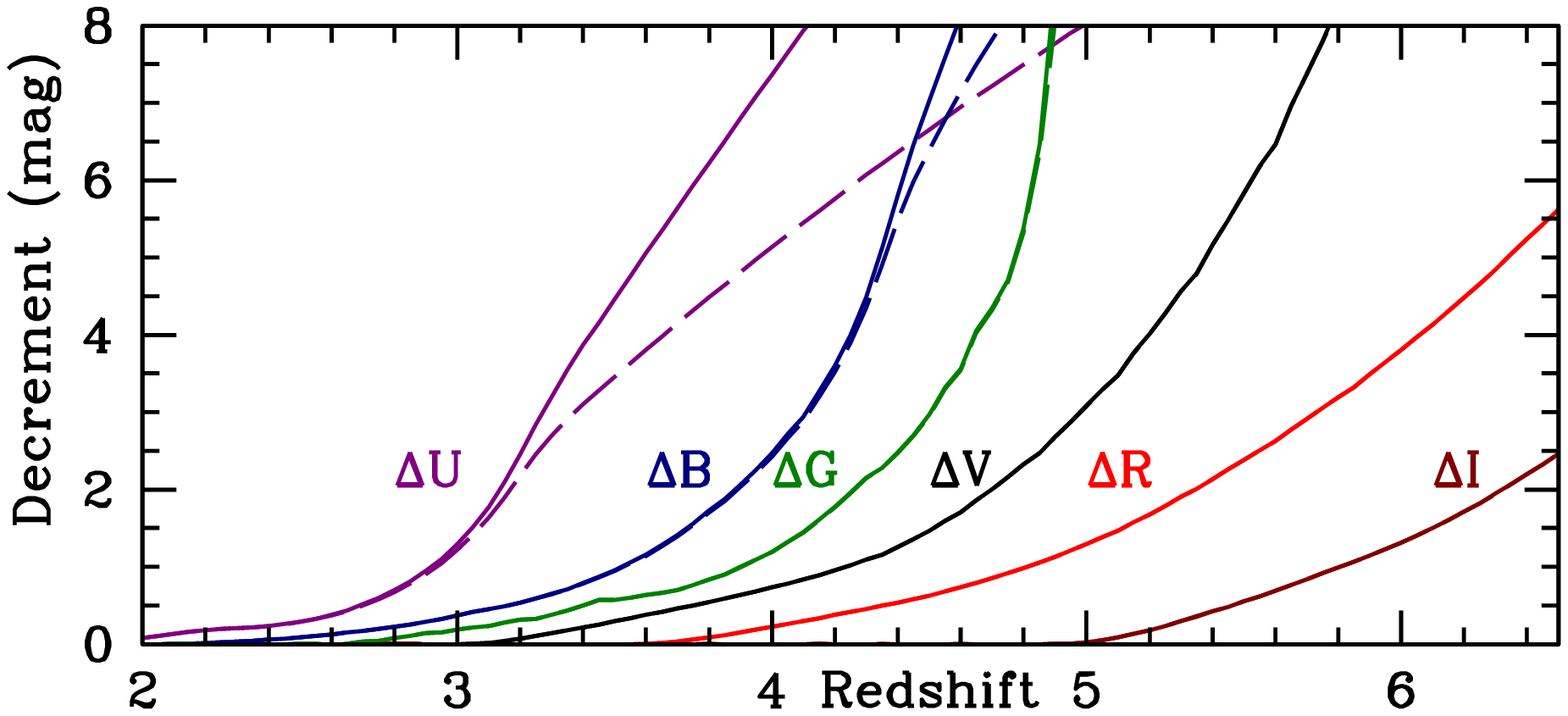, width=86mm, height=43mm}
\end{center}
\caption\protect{Variation of magnitude decrements with redshift for the
six broad-band filters used in our own observations.  Solid lines show the 
mean decrement due to {\em all}
absorbers, while dashed lines do not include the contribution of Lyman
limit systems.  Note in particular that the absence of LLS absorbers along
a particular line of sight is most likely to affect the chances of
detecting $U$ dropouts at redshifts $3 \ltabout z \ltabout 3.5$. Even so, a
substantial break should be observed even without a LLS absorber on the
line of sight.
\label{colfig:decrement}
}
\end{figure}

The chances of an object missing our selection cuts owing to the absence of
a LLS along the line of sight are greater for $U$ dropouts than for $B$
dropouts for two reasons.  Firstly, the relative contribution of the LLS to
the attenuation at $z\sim 3$ is greater than at $z\sim 4$ (see
Figure~\ref{colfig:decrement}).  Secondly, the expected number of LLS along
a line of sight increases significantly between $z\sim 3$ and $z\sim 4$.

\scite{moljak90} give values ranging from 0.25 at $z_e=2.5$ to about 0.05
at $z_e=4.5$ for the probability of encountering no LLS leading to an
absorption depth $\tau > \tau_{crit} = 1.5$ at the \HeII$\;$line $\lambda_e =
304$\AA.  As the photoionisation cross-section of H decreases approximately
as $\nu^{-3}$ below the Lyman limit, this optical depth corresponds to a
very optically-thick Lyman limit system, but these probabilities are a
useful indicator of how the likelihood of encountering a Lyman limit system
changes with redshift.

\section{Dust extinction}\label{sec:dustext}

While in the optical a measure of the dust extinction is obtainable from
the emission line Balmer optical depth $\tau^{l}_{B}$, dust extinction
in the UV is typically parametrized in terms of the continuum slope.
The UV continuum of a star-forming galaxy may be approximated by a power
law $F(\lambda) \propto \lambda^\beta$, with a ``colour excess'' defined
by $E(\beta) = \beta - \beta_0$.  Spectral synthesis models give a value
$-2.5 \la \beta_0 \la -2.0$, in broad agreement with measured values.  For a
young stellar population (age $\ltabout 2 \times 10^7$ yr), the
differences, other than of normalisation, between the spectra of the
extreme scenarios of instantaneous burst and constant star formation rate
are small and within measurement uncertainties.

We calculate dust extinction according to the method of \ncite{calzetti94}
(1994, hereafter C94), taking the extinction parameter $A(\lambda)$ to be
empirically fitted by a third-order polynomial $Q_e(x)$ (where
$x=1/\lambda$ such that $A(\lambda) = \tau{^l}{_B} Q_e(x)$.

$Q_e$ is purely the {\em selective} extinction as opposed to the {\em
total} extinction, since for any spectral slope $\beta$ one does not in
general know at what wavelength there is {\em no} extinction. 
To obtain the total extinction, we use the method described by
\scite{meurer95}, using consideration of energy conservation (relating the
far IR excess to $\beta$).
The ratio of total to selective extinction is parametrized as $X'_\lambda =
A_\lambda /E(\beta)$.  For the HST F220W filter (central wavelength
2280\AA, thus $Q_e({\mathrm {F220W}})=1.582$) and the C94 starbursts, this
value is constant at $X'_{F220W}=1.6$.  

Using the starburst values of C94 to give $\tau{^l}{_B}=0.494 E(\beta)$
and choosing $\beta_0 = -2.5$, we obtain

\begin{equation} 
f'(\lambda) = { 
        { 10^{-0.4 \times 1.6 (\beta + 2.5)} {f(\lambda)} \over
\exp\left[0.494 (\beta+2.5)\left(Q_e(\lambda) - 1.582\right)\right]  }}
\end{equation}

The dust extinction corrections are applied to our synthetic spectra
prior to correcting for intergalactic extinction, using values for $\beta$
of $-2.5$ (zero extinction), $-1.5$ and $-0.5$ (equivalent to $E(B-V)$ of
0.00, 0.23 and 0.46).

\section{Luminosity-SFR calibration}\label{sec:lumcalib}

\begin{figure}
\begin{center}
\epsfig{file=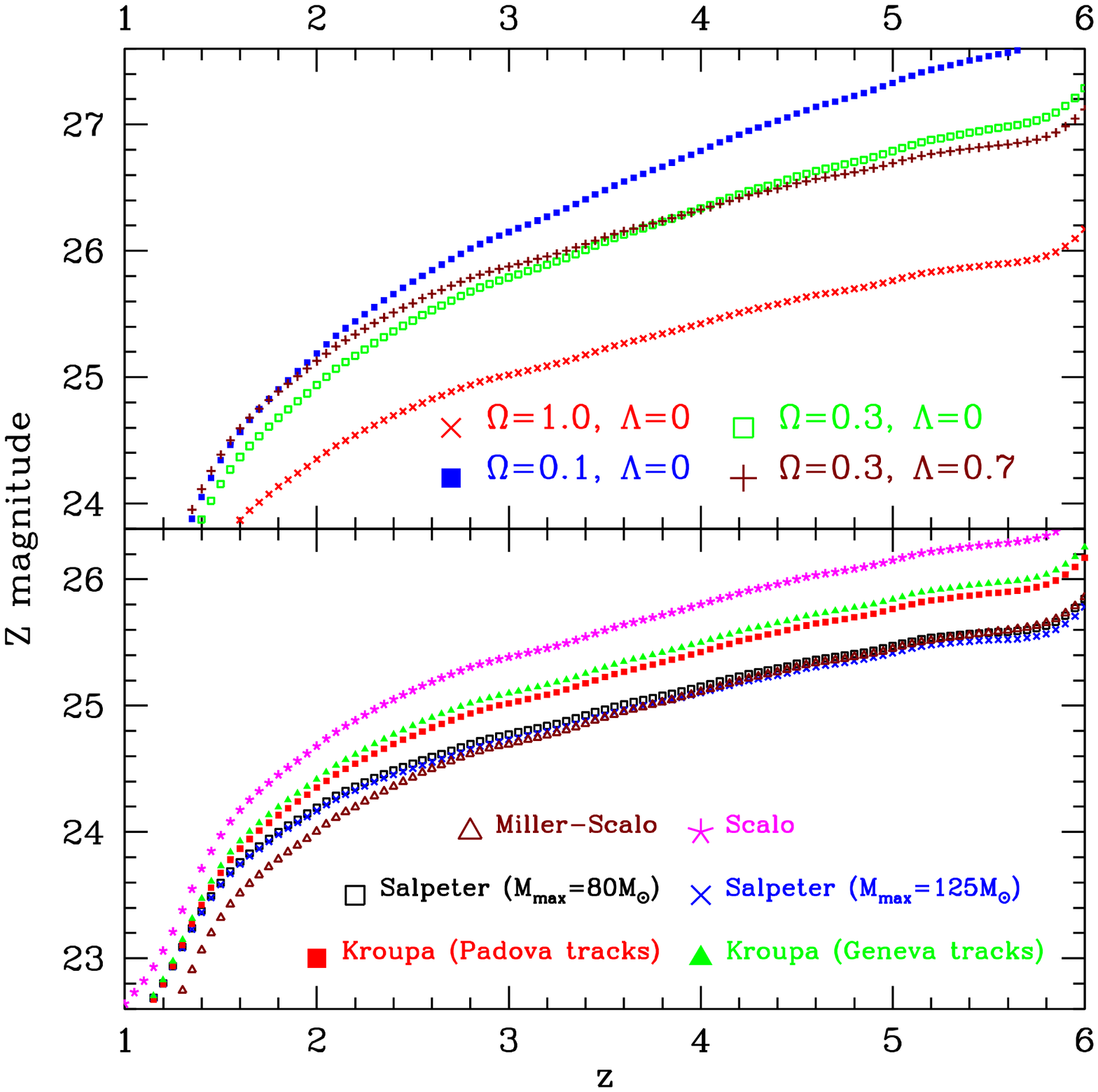, width=83mm, height=83mm}
\end{center}
\vspace{-4mm}
\caption\protect{
\label{colfig:sfrzcosmol}
Variation of $Z$ magnitude with redshift for an unreddened
star-forming galaxy with a continuous star formation rate of 10 $h_{50}^{-2}$
\Msun\ yr$^{-1}$.   The top plot shows the effect of changes in cosmology
for a Kroupa IMF.  The bottom plot shows the effect of varying the initial
mass function for $\Omega=1$, $\Lambda=0$ and Padova evolutionary tracks
except where indicated.  Sensitivity to the choice of evolutionary track is
at the 5\% level (though this is likely to increase if metallicity effects
are taken into account), but the choice of IMF has a very significant effect on
the luminosity-SFR calibration. 

$Z$ is used as \lya\ does not reach the band until $z\approx 5.8$, and
hence it represents the unattenuated UV continuum flux.  While the
unreddened UV continuum is frequently regarded as a flat spectrum, this is
merely an approximation, and $K$ correction effects lead to the variation
of magnitude with redshift differing slightly from the distance modulus.  
}
\end{figure}

For the calibration between observed magnitude and star-formation rate, we
choose for ease of comparison to use the same model as \scite{steidspec96},
namely a Salpeter IMF with upper mass cutoff at 80\Msun --- this low cutoff
has a significant effect on the magnitude (Figure~\ref{colfig:sfrzcosmol}).  

For an $\Omega=1$, $\Lambda=0$ cosmology with $\Ho=50$
km~s$^{-1}$~Mpc$^{-1}$, we find that a constant star formation rate of
10.0\Msun~yr$^{-1}$ with solar metallicity at $z=3.25$  corresponds to a
magnitude in $R$ of 24.88 at age 100Myr, tending asymptotically to
$R=24.79$ as the galaxy ages\footnote{By an age of 60Myr, the luminosity is
within 10\% of this figure, and within 1\% by an age of 400Myr.}.  This is
slightly fainter than the value quoted by \ncite{steidspec96}, but probably
insignificant in view of the use of different spectral synthesis codes and
stellar libraries.    

Figure~\ref{colfig:sfrzcosmol} shows the effect of changes in
IMF, cosmology and evolutionary tracks.  It is evident that accurate
determination of the shape of the Madau curve is dependent on the
cosmology, initial mass function and more accurate representations of
stellar evolution at low metallicities.

\section{Model colour-colour plots}\label{sec:modcolcol}

\begin{figure}
\begin{center}
\epsfig{file=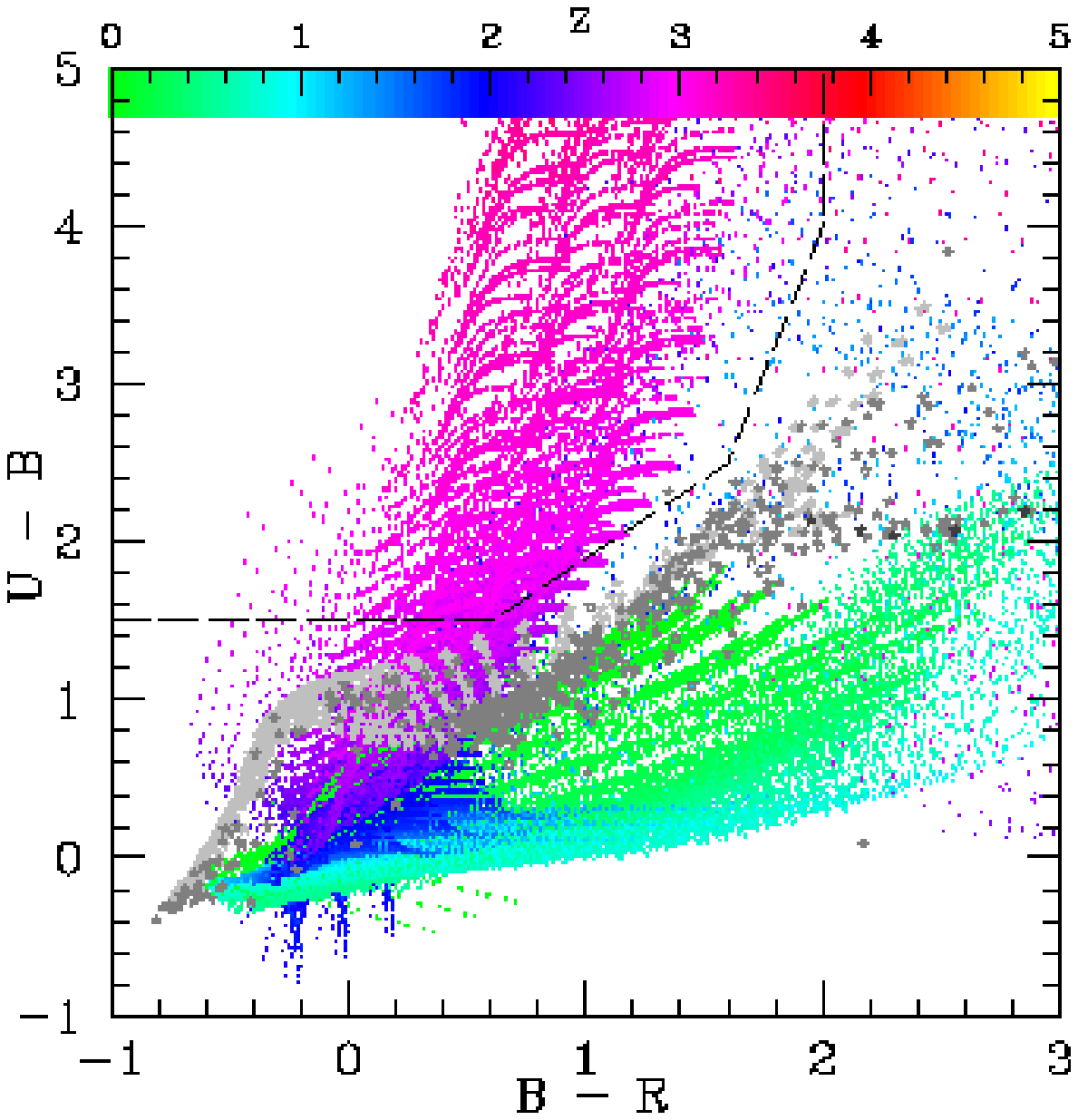, width=83mm, height=83mm}
\end{center}
\caption\protect{
Variation of $U-B$ and $B-R$ colours with redshift $z$.  Galaxies are
indicated by small dots coloured by redshift; young (age $<$100Myr)
galaxies in the redshift range $2.75 < z < 3.5$ have been emphasized.  The
colours of galactic stars have been superimposed: light grey stars are
those from the model data of \scite{bessel98}, while the darker grey stars
are from the photometric standards of \scite{landolt}.  Black stars
represent M dwarfs and use data obtained from \scite{mstars} and
\scite{leggett92}.

Our selection criteria are indicated by the dashed lines.  A broad area of
colour-colour space is enclosed, allowing for efficient selection of
candidate \hirs\ galaxies in spite of photometric errors, as long as a
sufficiently large break in $U-B$ can be observed.  In practice this as the
limiting factor, owing to the relatively poor quantum efficiency $U$ of the
TEK CCDs used compared to that of more modern detectors. 

Galaxies at redshifts above $z\approx3.5$ will have almost complete
absorption in $U$ and thus extreme $U-B$ colours off the top of this plot,
but in practice we will only measure lower limits to this colour.  As
redshift increases, increasing absorption in $B$ will push $B-R$ further
towards the red.  Making the upper bound to the cut in $B-R$ (or more
generally for the two longer wavelength bandpasses) further to the red
allows for selection of objects at higher redshifts than are shown on this
plot, however this must not be pushed so far as to reach the stellar locus.  
\label{colfig:ubbrsim}
}
\end{figure}

\begin{figure}
\begin{center}
\epsfig{file=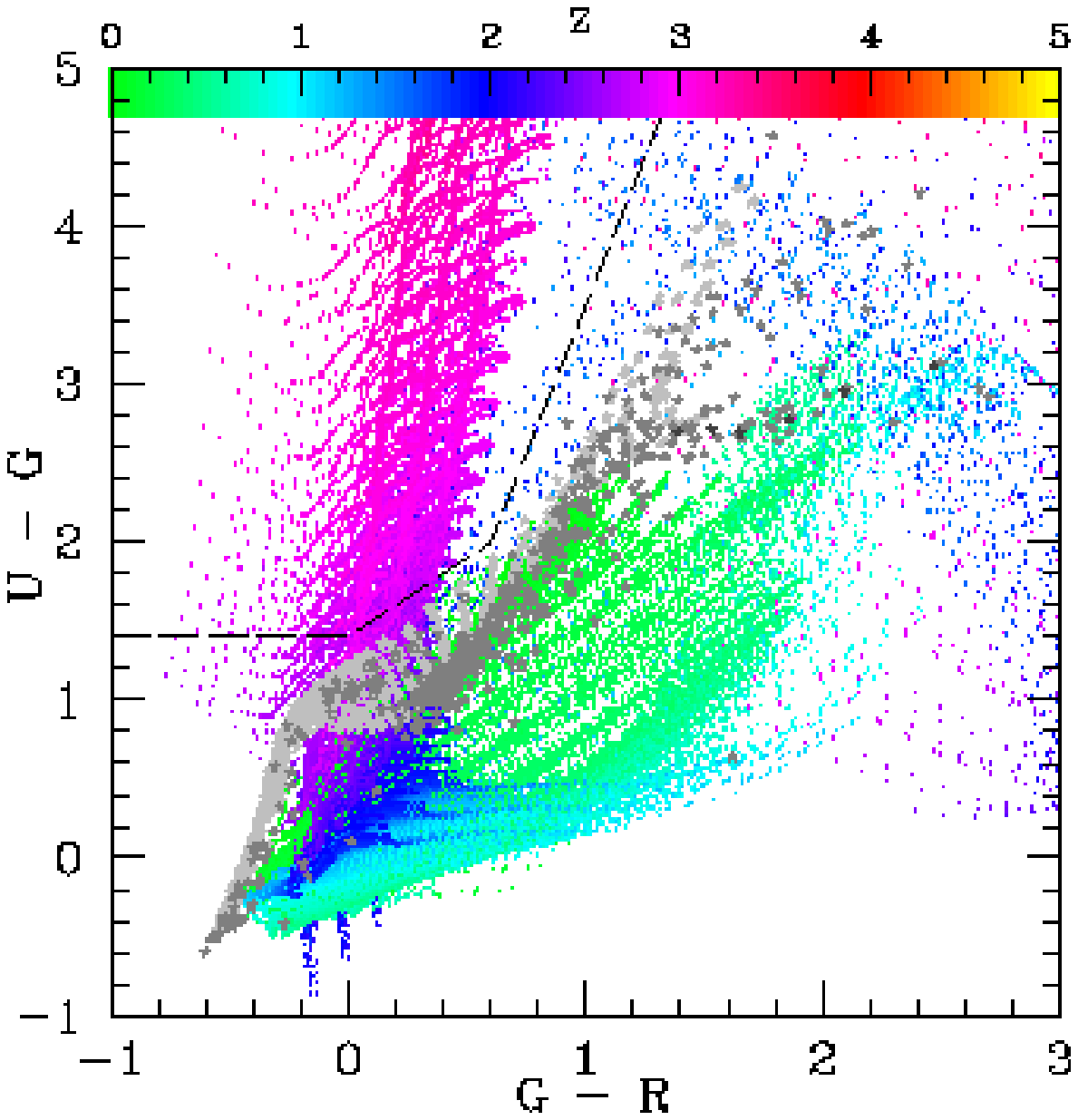, width=84mm, height=84mm}
\end{center}
\caption\protect{As Figure~\ref{colfig:ubbrsim} but for $U-G$ and $G-R$.  $G$
magnitudes for stars are derived by interpolation from the $B$ and $V$
magnitudes.  Note in particular the decrease in the area of colour-colour 
space in the selection area relative to Figure~\ref{colfig:ubbrsim}.
\label{colfig:uggrsim}
}
\end{figure}

\begin{figure}
\begin{center}
\epsfig{file=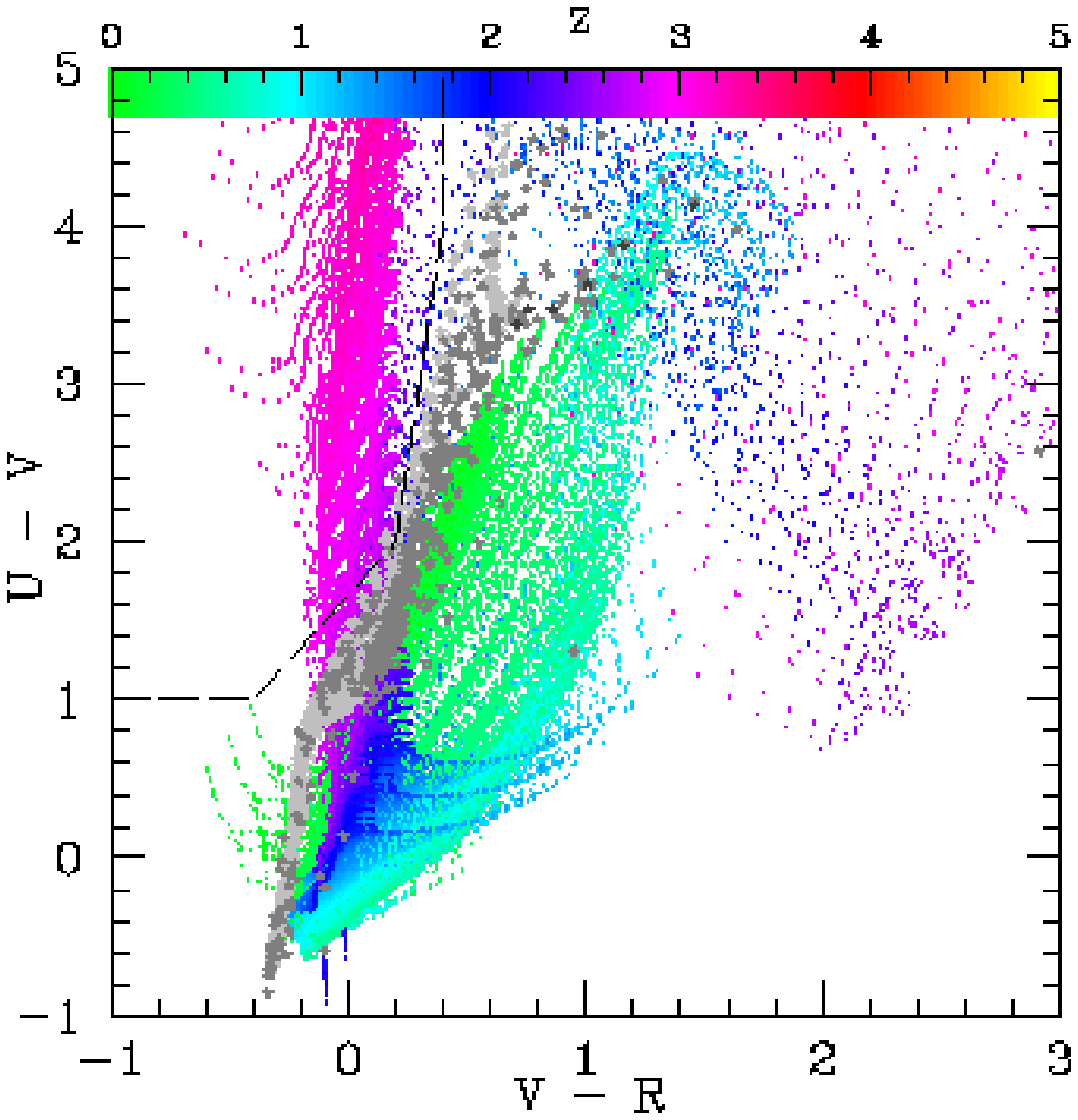, width=84mm, height=84mm}
\end{center}
\caption\protect{As Figure~\ref{colfig:ubbrsim} but for $U-V$ and $V-R$.  The
low $V-R$ colours for stars and low-redshift galaxies mean that the colour
must be tightly constrained to avoid severe problems with contamination of
candidate \hirs\ objects and use of these colours to select \hirs\
galaxies therefore not recommended.  However it
may provide a useful check in deep multicolour surveys.  
\label{colfig:uvvrsim}
}
\end{figure}

\begin{figure}
\begin{center}
\epsfig{file=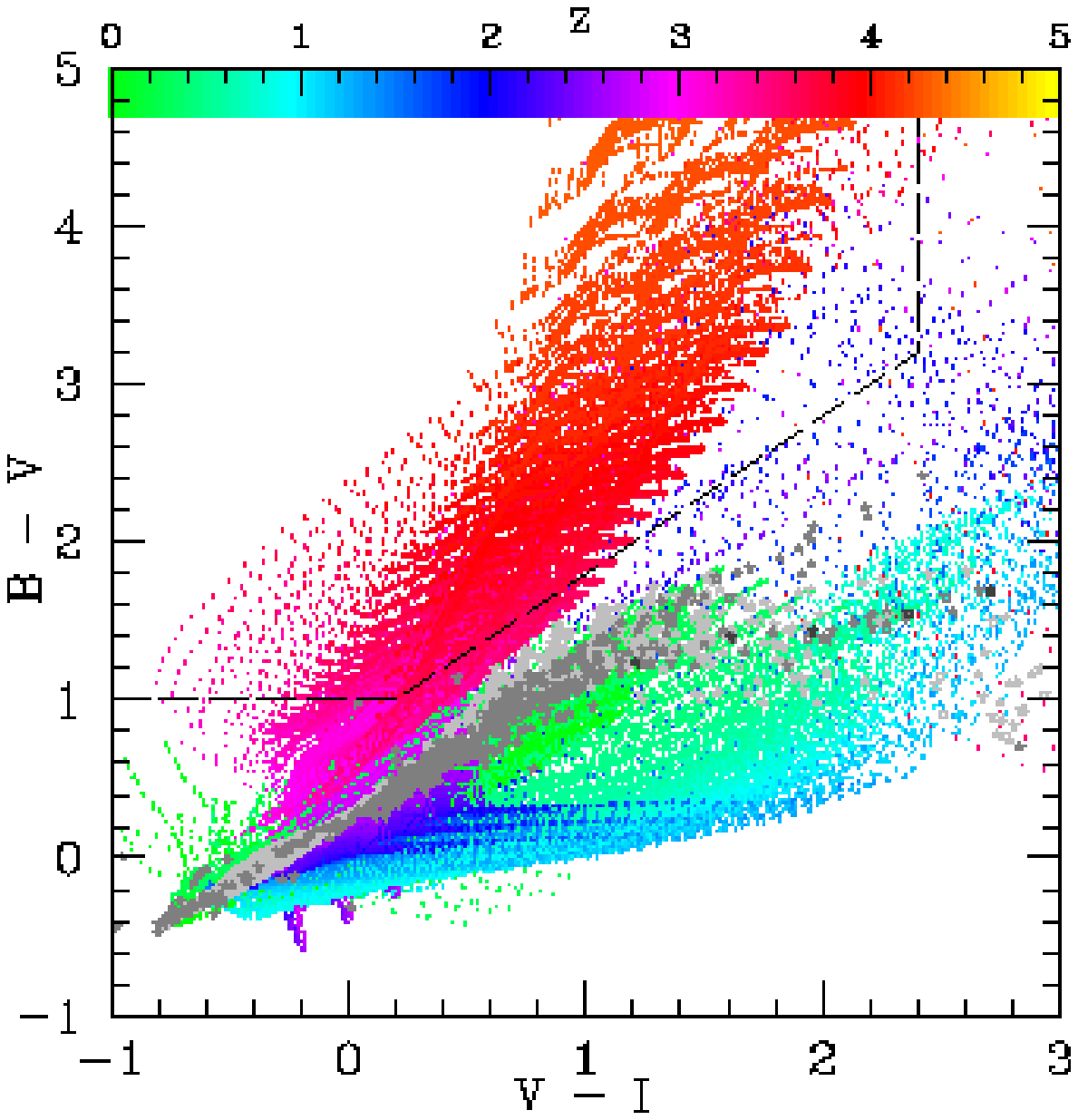, width=83mm, height=83mm}
\end{center}
\caption\protect{As Figure~\ref{colfig:ubbrsim} but for $B-V$ and $V-I$,
and with galaxies in the redshift range $3.5 < z < 4.5$ emphasized.  Again
there is good separation between the \hirs\ galaxy population and the low
redshift galaxies and stars.

\label{colfig:bvvisim}
}
\end{figure}

\begin{figure}
\begin{center}
\epsfig{file=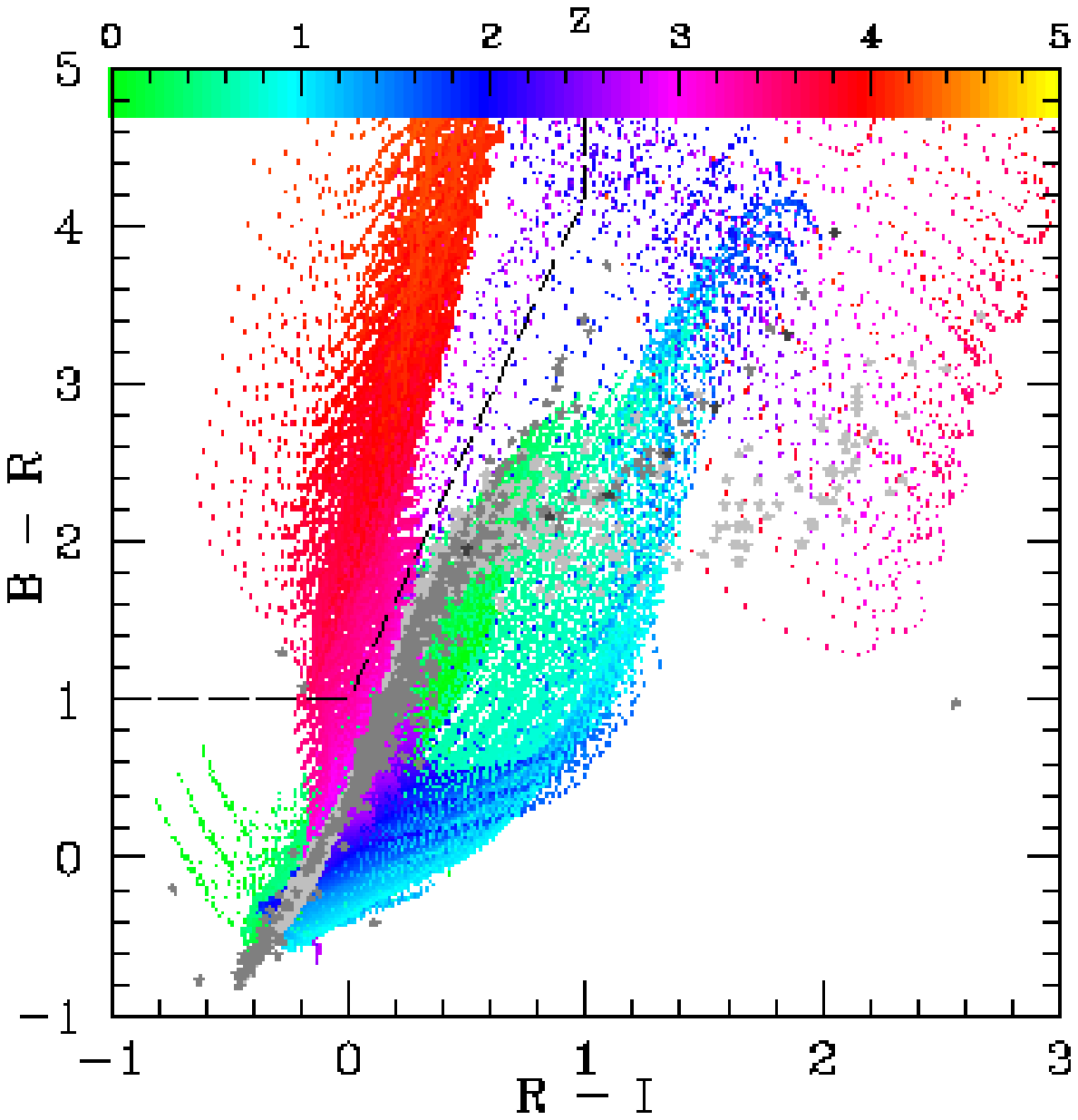, width=83mm, height=83mm}
\end{center}
\caption\protect{As Figure~\ref{colfig:bvvisim} but for $B-R$ and $R-I$.  
In comparison with the \textit{BVI} method (Figure~\ref{colfig:bvvisim}), the
separation between the \hirs\ galaxies and the stars and low redshift
galaxies is poor, making efficient selection harder without good limits on
the $B-R$ colour.
\label{colfig:brrisim}
}
\end{figure}

\begin{figure}
\begin{center}
\epsfig{file=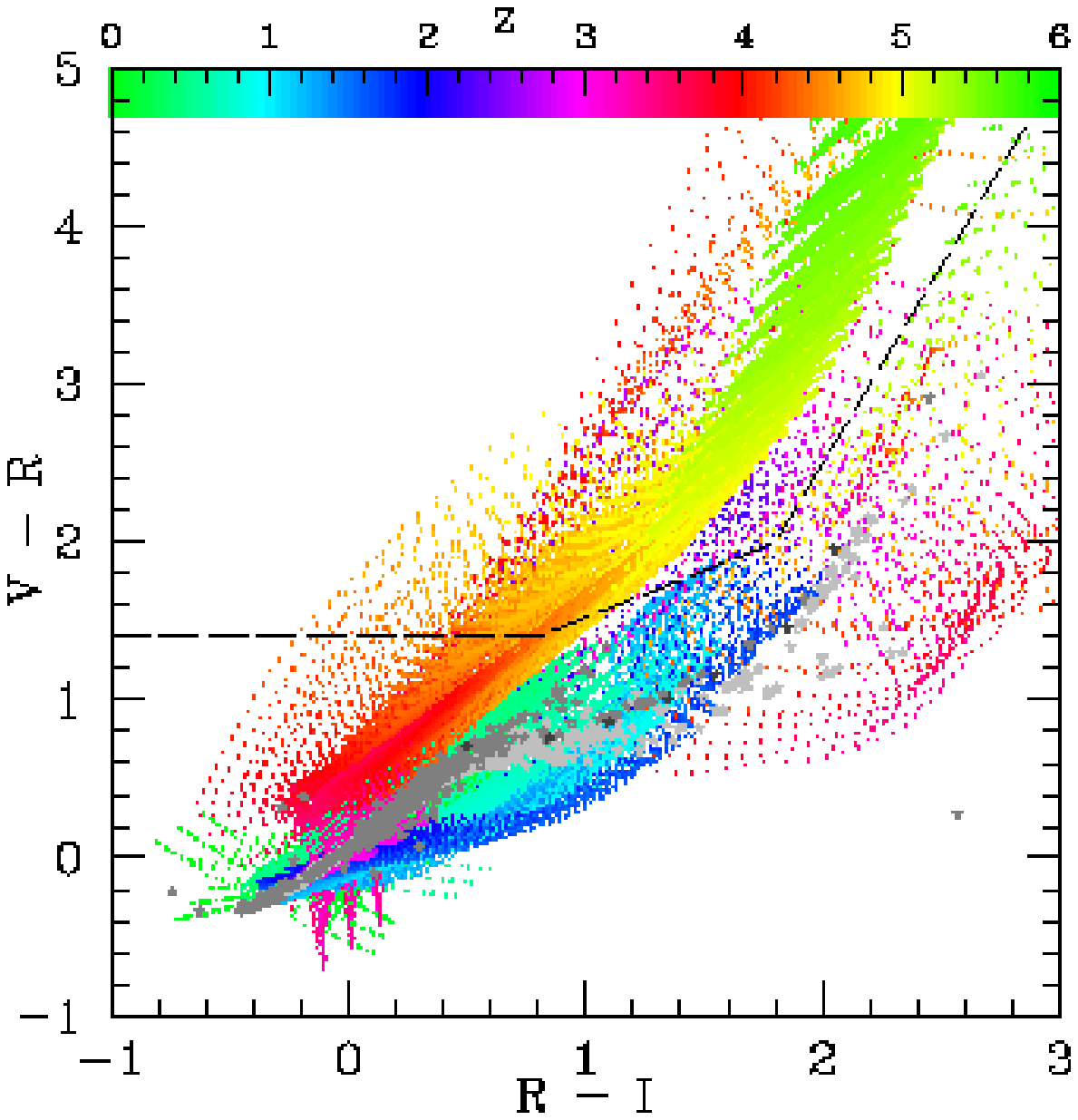, width=83mm, height=83mm}
\end{center}
\caption\protect{As Figure~\ref{colfig:brrisim} but for $V-R$ and $R-I$.  
Selection of candidates at $z \sim 5$ is perfectly feasible, but
spectroscopic confirmation of objects becomes increasingly difficult as 
the redshift increases.
\label{colfig:vrrisim}
}
\end{figure}

Colour-colour plots for some of the three-filter photometric systems used 
in our observations are presented in Figures~\ref{colfig:ubbrsim} to
\ref{colfig:vrrisim}.  A wider range of plots is available via the WWW at
our website \cite{rejsweb},
together with an interface for the generation of plots for arbitrary colour
pairs within a substantial database of optical, NIR and HST filters.

Galaxy colours are plotted at redshift increments of 0.05, together with
colours typical of galactic stars, and our chosen selection criteria are
shown.  Young galaxies within the targetted redshift ranges have been
highlighted for clarity.  

In producing these plots, no assumptions have been made as to whether
objects are likely to be observable.  Galaxies are shown at all redshift
intervals for which the ages of the oldest stars are less than the age of
the Universe in any reasonable cosmology.
In some cases evolved \hirs\ objects will therefore be shown which are
unphysically old or at least exceedingly unlikely to be observable within
the foreseeable future.

Such plots can be used to determine the most effective use of broad-band
filters to detect Lyman-break objects in future surveys.  For efficient
selection one must try to separate the locus of Lyman-break objects as much
as possible from that of stars and low-redshift galaxies, an important
factor is thus to ensure a reasonable span in wavelength between the
filters used for selection.  Figure~\ref{colfig:uvvrsim} clearly shows the
problem in using $V$ and $R$ as the two longer-wavelength bandpasses, since
the two are too close in effective wavelength to give a significant colour
term on intrinsically red objects.  Using $G$ as the central filter is
better (Figure~\ref{colfig:uggrsim}) but Figure~\ref{colfig:ubbrsim} shows
that $B$ is optimum.  

Obviously the choice of filters will be influenced somewhat by the
targetted redshift, three-filter selection working most effectively where
the central filter spans the wavelength of redshifted \lya.  \textit{UBR}
will be ineffective at selection of objects much beyond $z\approx3.1$,
whereas Steidel and co-workers have already shown the effectiveness of a
system similar to our \textit{UGR} out to around $z\approx3.5$.  Again,
\textit{UGR} is inappropriate beyond this: for instance we note that the
\rg\ \fourc\ at $z=3.80$ is too red in $G-R$ to be selected in
\textit{UGR}, but is a clear dropout in \textit{UVR}.

When searching for objects beyond $z\approx3.5$, it becomes preferable to
use $B$ rather than $U$ as the shortest wavelength bandpass.  While
\textit{UVI} selection would in principle work well, the increased
observing time required to make sufficiently deep observations in $I$
rather than $R$ owing to the increased sky background makes it advantageous
to use $B$ in place of $U$ to decrease the necessary observing time in the
bluest filter.  As a central wavelength, $V$ is preferable to $R$ owing to
the greater separation between high-redshift galaxies and low redshift
objects, as can be seen from Figures~\ref{colfig:bvvisim} and
\ref{colfig:brrisim}.

\section{Extensions to $z > 4.5$}
Selection of candidates at redshifts beyond $z\sim4.5$ is perfectly
feasible given sufficiently deep observations, although spectroscopic
confirmation of such objects becomes much more difficult due to absorption
and emission lines becoming redshifted to among the OH bands of the
near-IR.  

Steidel's group have added an $I$ filter to their $\steidU\steidG\steidR$
system in order to look for $\steidG$ dropouts at $z\sim4.5$, and have had
considerable success \cite{steidg99}.  At higher redshifts still, $V$
dropouts can be searched for.  A \textit{VRZ} filter system is in theory
very effective at selecting galaxies at $z\sim5$, but observational
considerations will in general make this impractical owing to the very high
sky background, poor sensitivity of CCDs and severe problems with fringing
if thinned CCDs are used.  

In practice \textit{VRI} is likely to be more suitable, indeed we use the
data obtained in one of our fields to conduct a pilot study, but conclude
that our data are insufficiently deep to probe a population of $V$ dropouts
at $z\sim5$.  $Z$ imaging could be more effective with the use of
large-format infra-red arrays, which are a factor of $\sim2$ more sensitive
than CCDs at such wavelengths, but the much greater area of CCDs,
particularly in the form of large mosaic cameras, will give them the edge
for some time to come.

Deep multicolour imaging surveys using large-format detectors will reveal
many candidates.  One such survey is being undertaken by researchers at
Oxford using the Wide Field Camera
on the Isaac Newton Telescope, imaging ten square degrees in $BVRI$ to
depths comparable with this work, with subareas also imaged in $U$ and $Z$.
Among the many possibilities of these data is that of searching for
bright Lyman-break galaxies spanning the range $2.8 \la z \la 5$.
Given the detection limits of the survey, we expect to find $\sim15000$ $U$
dropouts at $z\sim3$, 2000 or more $R$ dropouts at $z\sim4$ and $\sim100$
$V$ dropouts at $z\sim5$ \cite{rejsthesis}.

Shallower surveys can still make use of similar techniques for finding
\hirs\ objects.  Though insufficiently deep to find galaxies at such
redshifts, the Sloan Digital Sky Survey will probe quasars and other bright
active galactic nuclei. Very similar colour selection techniques to ours
have been applied to commissioning data to identify candidate quasars at
redshifts above $z\ga 3.6$ \cite{sdss-quasars99,sdss-quasars00}.

Beyond $z \sim 6$, intergalactic attenuation will affect even $Z$ band and
objects at $z>6$ with star-formation rates typical of
objects at $z\sim3$ will have NIR magnitudes of 26 or beyond, difficult to
observe even with long observations on 8--10m class telescopes.
Ultimately observations of the earliest star-formation in the Universe
at $10 \la z \la 30$ must be the preserve of NGST and similar instruments. 

\begin{table}
\begin{center}
\begin{minipage}{0.9\linewidth}
\begin{center}
\begin{tabular}{lccc}
Field & $z$ & \multicolumn{2}{l}{Filter choice} \\
&  & Ideal & of \textit{UBVRIZ} \\
\hline
? & $\ga5.3$ & $r'i'z'$ &\textit{RIZ} \\
TN\,J0924-2201 & 5.19 & \textit{VRI} & \textit{VRI}\\
\sixc\ & 4.41 & \textit{GRI}  & \textit{BVI} \\
\eightc\ & 4.25 & \textit{GRI}  & \textit{BVI} \\
\fourc\ & 3.80 & \textit{BVI}  & \textit{BVI} \\
6C\,0032+412 & 3.65 & \textit{BVI}  & \textit{BVI} \\
\btwo\ & 3.40 & \textit{UGR} & \textit{UBR} 
\\
& $\la 3.0$ & $u'$\textit{GR} & \textit{UBR} \\
\hline
\end{tabular}
\end{center}
\label{tab:idealfilters}
\protect\caption{Choice of filters for searching for galaxies at 
similar redshift to the central \rg\ in various \hirs\ \rg\ fields,
including those targetted in our own work \protect\cite{rejspaper}.
$u'g'r'i'z'$ refer to the SDSS filter set.  At $z\la3$ a filter slightly
bluer than the standard $U$ filter such as $u'$ or the \steidU\ of Steidel
\etal\ will give a sharper break below \lya, while the lesser overlap
between the filters of the Sloan set are more suitable when searching for
$R$ dropouts.  Inevitably the choice depends on the availability of filters
for a particular instrument, but in many cases use of a customized filter
set, with minimal overlap between filter bandpasses, will provide greater
efficiency at a particular target redshift. In particular we conclude that
such techniques are likely  to perform very poorly for target redshifts 
$4\la z \la 4.7$ if one is restricted to a standard set without a 
\textit{G} or similar filter between \textit{B} and \textit{V}.  }
\end{minipage}
\end{center}
\end{table}

\section{Acknowledgements}

We thank Steve Rawlings, Lance Miller, Max Pettini, Gavin Dalton 
and Tony Lynas-Gray for helpful discussions.
\ackpparc\ackads

\bibliographystyle{newmnras}
\bibliography{acronym,common}

\bsp

\end{document}